\documentclass[final,3p,times,twocolumn]{elsarticle}
\usepackage{lineno,hyperref}
\usepackage{array}
\usepackage{makecell}
\usepackage{epstopdf}
\usepackage{amsmath,amsfonts,amsthm}
\usepackage{graphicx} 
\usepackage{float}

\usepackage{stfloats}

\usepackage{amssymb}
\usepackage[FIGBOTCAP]{subfigure}
\usepackage{upgreek}
\usepackage{multirow}
\usepackage{float}

\usepackage{multirow}

\usepackage{textcomp,booktabs}
\usepackage[usenames,dvipsnames]{color}
\usepackage{colortbl}

\journal {Nuclear Instruments and Methods in Physics Research Section A}

\begin{document}
\begin{frontmatter}
	
	\title{1.28 and 5.12 Gbps multi-channel twinax cable receiver ASICs for the ATLAS Inner Tracker Pixel Detector Upgrade}
	
	\author[mymainaddress,mysecondaryaddress]{Chufeng Chen}
	
	\author[mymainaddress]{Datao Gong\corref{mycorrespondingauthor}}
	\cortext[mycorrespondingauthor]{Corresponding author}
	\ead{dgong@mail.smu.edu}
	
	\author[mythirdaddress]{Suen Hou}
	\author[mysecondaryaddress]{Guangming Huang}
	\author[mymainaddress,mysecondaryaddress]{Xing Huang}
	\author[myfourthaddress]{Szymon Kulis}
	\author[myfifthaddress]{Paul Leroux}
    \author[mymainaddress]{Chonghan Liu}
    \author[mymainaddress]{Tiankuan Liu}
    \author[myfourthaddress]{Paulo Moreira}
    \author[myfifthaddress]{Jeffery Prinzie}
    \author[mymainaddress]{Peilong Wang}
    \author[mymainaddress]{Jingbo Ye}

	\address[mymainaddress]{Department of Physics, Southern Methodist University, Dallas, TX 75275, USA	}
	\address[mysecondaryaddress]{Department of Physics, Central China Normal University, Wuhan, Hubei 430079, P.R. China}
    \address[mythirdaddress]{Institute of Physics, Academia Sinica, Taiwan}
    \address[myfourthaddress]{Microelectronic group at CERN, 1211 Geneva 23, Switzerland}
    \address[myfifthaddress]{KU Leuven, Leuven, Belgium}
	
	\begin{abstract}
        We present two prototypes of a gigabit transceiver ASIC, GBCR1 and GBCR2, both designed in a 65-nm CMOS technology for the ATLAS Inner Tracker Pixel Detector readout upgrade.

        The first prototype, GBCR1, has four upstream receiver channels and one downstream transmitter channel with pre-emphasis. Each upstream channel receives the data at 5.12 Gbps through a 5 meter AWG34 Twinax cable from an ASIC driver located on the pixel module and restores the signal from the high frequency loss due to the low mass cable. The signal is retimed by a recovered clock before it is sent to the optical transmitter VTRx+. The downstream driver is designed to transmit the 2.56 Gbps signal from lpGBT to the electronics on the pixel module over the same cable. The peak-peak jitter (throughout the paper jitter is always peak-peak unless specified) of the restored signal is 35.4 ps at the output of GBCR1, and 138 ps for the downstream channel at the cable ends. GBCR1 consumes 318 mW and is tested.

        The second prototype, GBCR2, has seven upstream channels and two downstream channels. Each upstream channel works at 1.28 Gbps to recover the data directly from the RD53B ASIC through a 1 meter custom FLEX cable followed by a 6 meter AWG34 Twinax cable. The equalized signal of each upstream channel is retimed by an input 1.28 GHz phase programmable clock. Compared with the signal at the FLEX input, the additional jitter of the equalized signal is about 80 ps when the retiming logic is off. When the retiming logic is on, the jitter is 50 ps at GBCR2 output, assuming the 1.28 GHz retiming clock is from lpGBT. The downstream is designed to transmit the 160 Mbps signal from lpGBT through the same cable connection to RD53B and the jitter is about 157 ps at the cable ends. GBCR2 consumes about 150 mW when the retiming logic is off. This design was submitted in November 2019.

	\end{abstract}


\end{frontmatter}	



\section{Introduction}
   In the Phase-II upgrade of the ATLAS inner tracker (ITk) Pixel Detector, the pixel sensors are to be read out by the RD53B ASIC and its output is to be transmitted to a location meters away from the sensors where it is to be converted to optical for further transmission through optical fibers \cite{Ref1}. The output data rate of RD53B is 1.28 Gbps. The readout scheme evolves from a data aggregator ASIC on the pixel module to send 4*1.28 = 5.12 Gbps signal through a 5 meter AWG34 Twinax cable to routing the 1.28 Gbps RD53B output out through a 1 meter custom FLEX cable followed by a 6 meter AWG34 Twinax cable and to use lpGBT as the data aggregator. In both cases VTRx+ will be used to convert between electrical and optical signals. The latter becomes the baseline of the sensor readout, taking into consideration of ASIC development status, detector material budget, powering and implementation. In this baseline design to place the opto-box (which houses lpGBT, VTRx+ and other electronics) meters away from the sensor is also to avoid the high radiation environment near the interaction point and to offer relatively easy access to service and repairing. In both schemes one of the challenges is to transmit giga-bit persecond data over low-mass electrical cables. In the chosen AWG34 Twinax cable with a length from 3 to 6 meters the high-frequency loss is significant and the eye-diagram is fully closed even at 1.28 Gbps. To compensate this loss, an equalizer ASIC is needed at the cable ends. The peaking strength needs to be programmable cope with the system implementation where cable length varies. Although the opto-box is meters away from the interaction point, it is still in radiation environment. Beside the radiation-tolerant requirement, the power budget of the equalizer chip is also tight which makes the design more challenging.
   
   We present two prototype chips in this paper, both used to equalize the high-speed data signals over the low-mass cables. GBCR1 has 4 upstream channels each working at 5.12 Gbps, corresponding to the initial data transmission scheme \cite{Ref1}. The chip has been successfully tested at laboratory and in irradiation. GBCR2 is for the revised readout scheme which is now the baseline. GBCR2 has 7 upstream channels each at 1.28 Gbps, resulting tighter power consumption of each channel. Both ASICs have downstream cable driver channels to send signals to the electronics on the pixel module. 

\section{Design and Test for GBCR1}
\begin{figure}[h]
        \centering
        \setlength{\abovecaptionskip}{0pt}
        \includegraphics[width=1\linewidth]{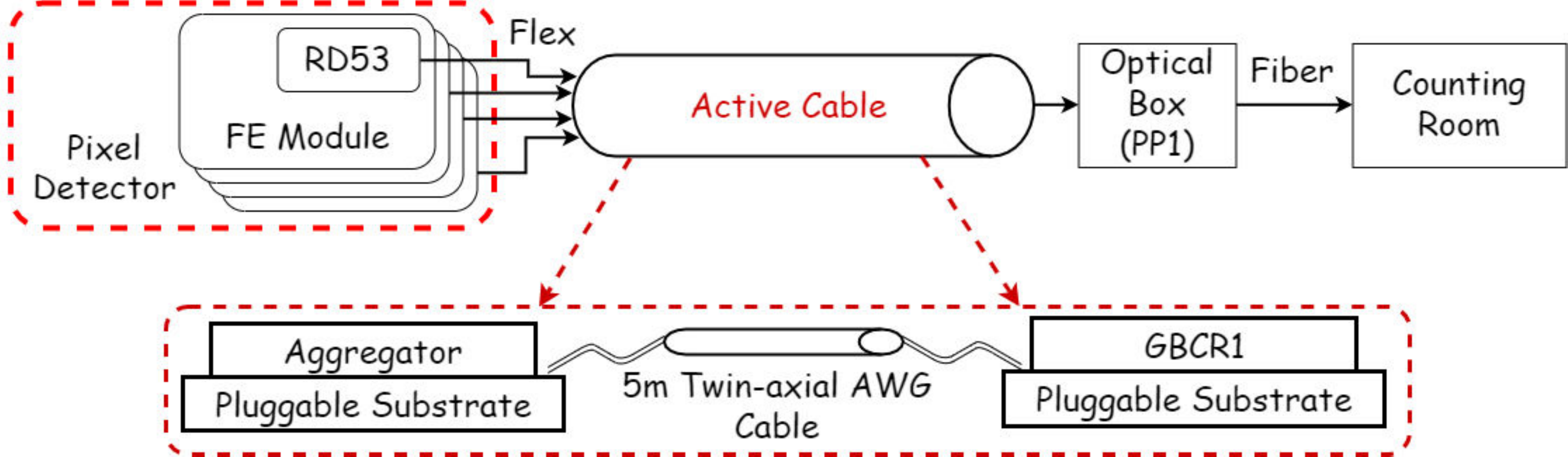}
        \caption{The original ITk Readout System Plan}
        \label{fig1:System_Plan}
    \end{figure}

    The initial scheme of data transmission between the pixel module and opto-box is based on a concept of "active cable" in which two dedicated ASICs, the Aggregator and GBCR1 are at the cable ends. Figure \ref{fig1:System_Plan} shows a diagram of this scheme. The Aggregator combines the four 1.28 Gbps data lanes from the FE modules \cite{Ref2} and transmits data at 5.12 Gbps to GBCR1 through a 5 meter AWG34 Twinax cable. GBCR1 receives these signals and recovers them to meet the requirements of the optical transmitter VTRx+ \cite{Ref3} \cite{Ref4} in the opto-box located at the patch-panel 1 (PP1). With the high-frequency loss compensated by GBCR1, the active cable performs as a wide-band cable. 

\begin{figure}[h]
    \centering
    \setlength{\abovecaptionskip}{0pt}
    \includegraphics[width=1\linewidth]{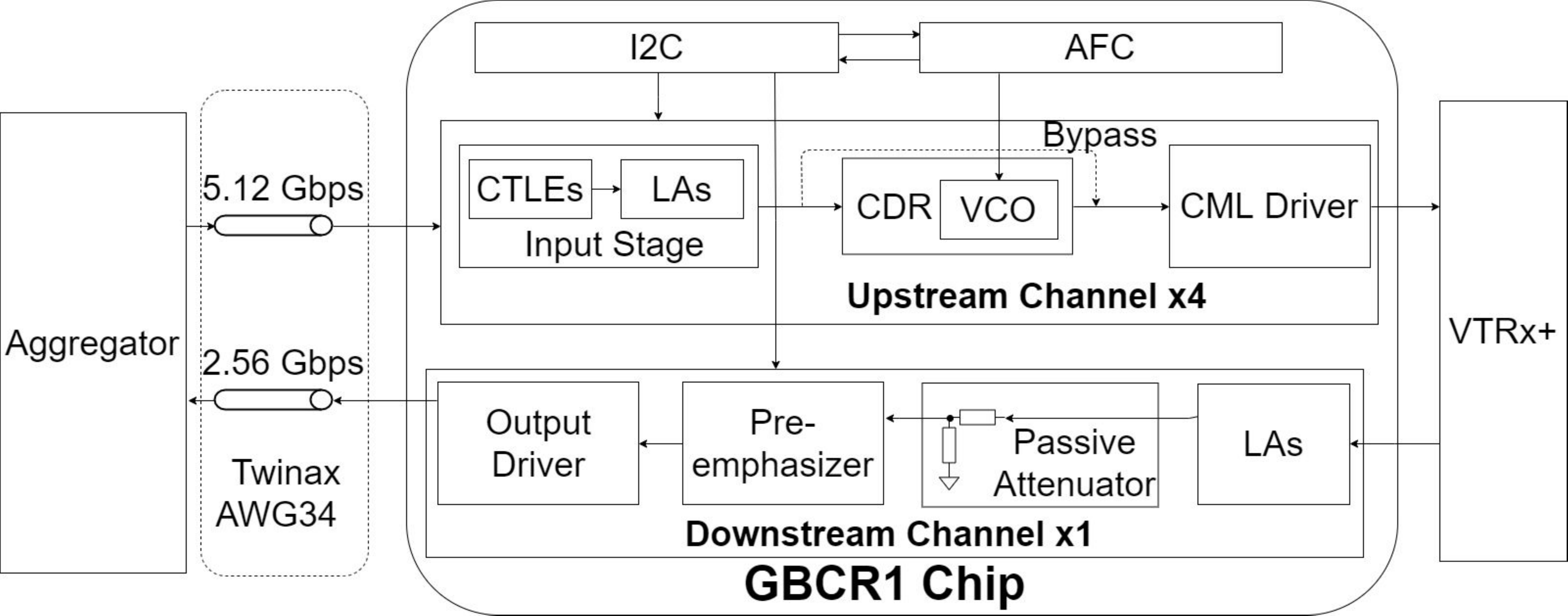}
    \caption{Block Diagram of GBCR1}
    \label{fig2:GBCR1}
    \end{figure}

   The block diagram of GBCR1 is shown in Figure \ref{fig2:GBCR1}. There are four upstream channels, one downstream channel, an Inter-Integrated Circuit (I2C) slave and an Automatic Frequency Calibration (AFC) module in GBCR1. Among these four upstream channels, there are 3 baseline channels and 1 test channel which is not presented in this paper. Each upstream channel operates at 5.12 Gbps data and consists of a Continuous Time Linear Equalizer (CTLE), a Limiting Amplifier (LA), a Clock Data Recovery (CDR) and a Current Mode Logic (CML) driver. The downstream channel includes 3 stages of LAs, a passive attenuator, a pre-emphasizer and an output driver with data rate at 2.56 Gbps. The AFC module \cite{Ref5} is used to calibrate the VCO in each upstream channel to ensure it works perfectly at 5.12 GHz \cite{Ref6}. The I2C slave provides all configuring and control signals of the whole chip.

    \subsection{Upstream Channel}
        The equalizer used in GBCR1 is based on a CTLE structure. The passive resistor and capacitor create a zero and a pole in the transfer function to compensate for channel loss \cite{Ref7}. CTLE behaves as a high-pass filter to compensate the undesired low-pass effect of the channel at high frequency. The active CTLE with RC can provide high-frequency gain peaking by means of a real zero \cite{Ref8}. The zero and pole are described by the following formulae:  
        
        \begin{equation}
            \omega_{Z} = \frac{1}{R_{S}C_{S}}
        \end{equation}
        \begin{equation}
            \omega_{p1} = \frac{1+\frac{g_{m}R_{S}}{2}}{ R_{S}C_{S}}
        \end{equation}
        
        The equalization strength is determined by the ratio of the zero and the pole, as described in formula
         \ref{con:EQ_Strength},
         
        \begin{equation}
            Equalization Strength =\frac{\omega_{Z}}{\omega_{p1}}
            = 1+\frac{R_{S}C_{S}}{2}\label{con:EQ_Strength}
        \end{equation}

\begin{figure}[h]
            \centering
            \setlength{\abovecaptionskip}{0pt}
            \includegraphics[width=1\linewidth]{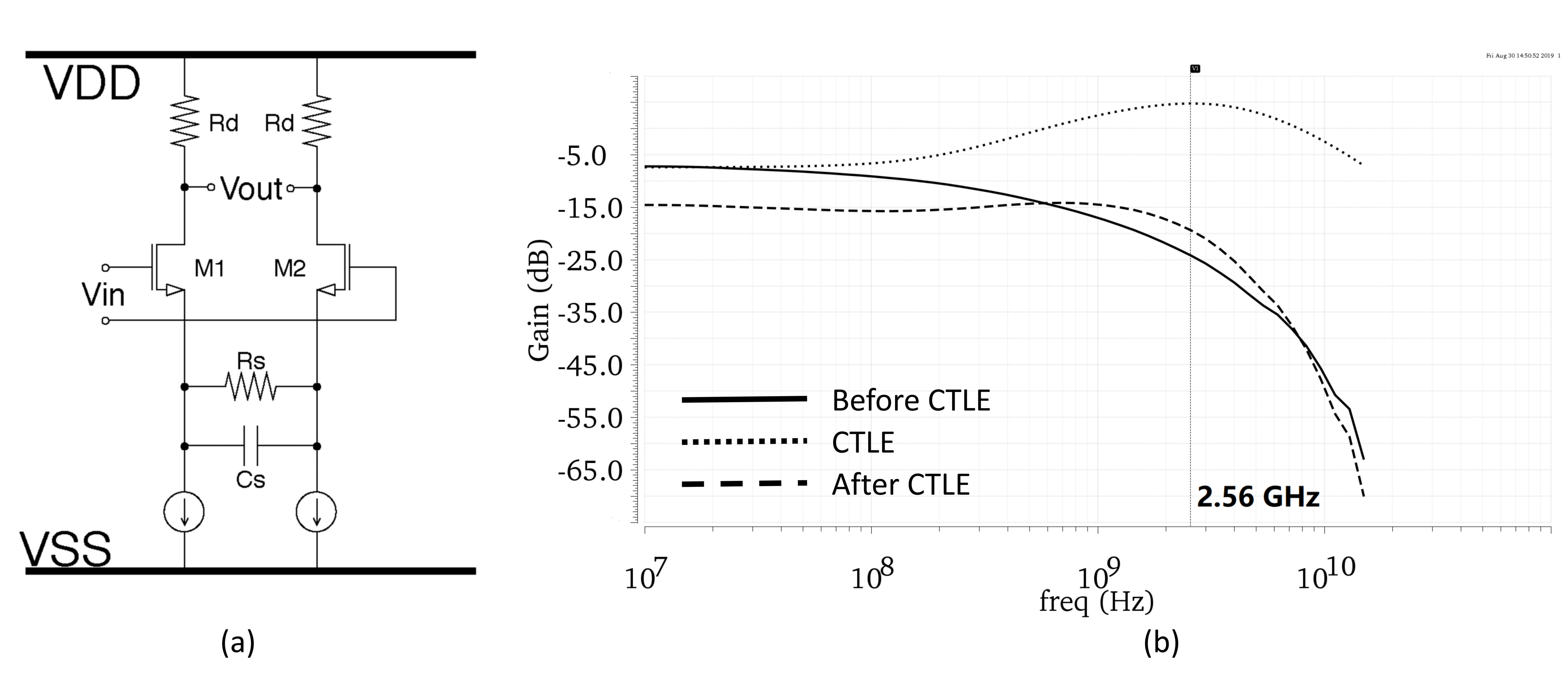}
            \caption{Schematic of a CTLE (a) and AC response of the equalizer (b)}
            \label{fig3:CTLE_and_equalizer}
        \end{figure}

       In our design, the $R_S$ and $C_S$ in the CTLE are programmable in order to adjust the zero, pole and equalization for the optimal results in different situations. As the solid line in Figure \ref{fig3:CTLE_and_equalizer}(b) shows, the signal loss at the Nyquist frequency, 2.56 GHz, of the cable is about -16.5 dB . The dotted line shows the frequency response of the equalizer with the peaking strength of 16.9 dB. After the equalization, the overall frequency response is flat as shown with the dash line. As the multiple CTLE stages create multiple order poles above 2.56 GHz, the frequency response curve falls rapidly beyond this frequency. Because the signal is saturated after the last stage of output buffer which bandwidth is 4.2 GHz, much higher than 2.56 GHz, the output signal rise and fall time is about 40 ps, much faster than the single-frequency of the Nyquist frequency.
       
       To further reduce the ISI jitter due to the cable, a CDR is used to recover the clock in the signal and retime the output. The CDR \cite{Ref9} in GBCR1 is adapted from lpGBT \cite{Ref10}. The original CDR works for 2.56 Gbps data only and the VCO is locked at 5.12 GHz. We remove the frequency detector to release the frequency feedback in the loop control so that it also works at 5.12 Gbps data rate \cite{Ref11}. The LC-VCO, the core of the CDR, has low phase noise but its oscillating range is narrow. A programmable capacitor bank is used to extend its oscillating range. The AFC is needed to select the optimal setting for the capacitor bank to ensure the operation at 5.12 GHz. 
       
       CML driver \cite{Ref12}, a differential amplifier with a load resistance of 50 Ohm, with a bandwidth of 4.2 GHz and an amplitude larger than 200 mV to meet the eRx requirements of lpGBT which is 140 mV, is used to transmit the recovered signals and drive the VTRx+ optical module in the next stage.

    \subsection{Downstream Channel}
\begin{figure}[h]
        	\centering
        	\setlength{\abovecaptionskip}{0pt}
        	\includegraphics[width=1\linewidth]{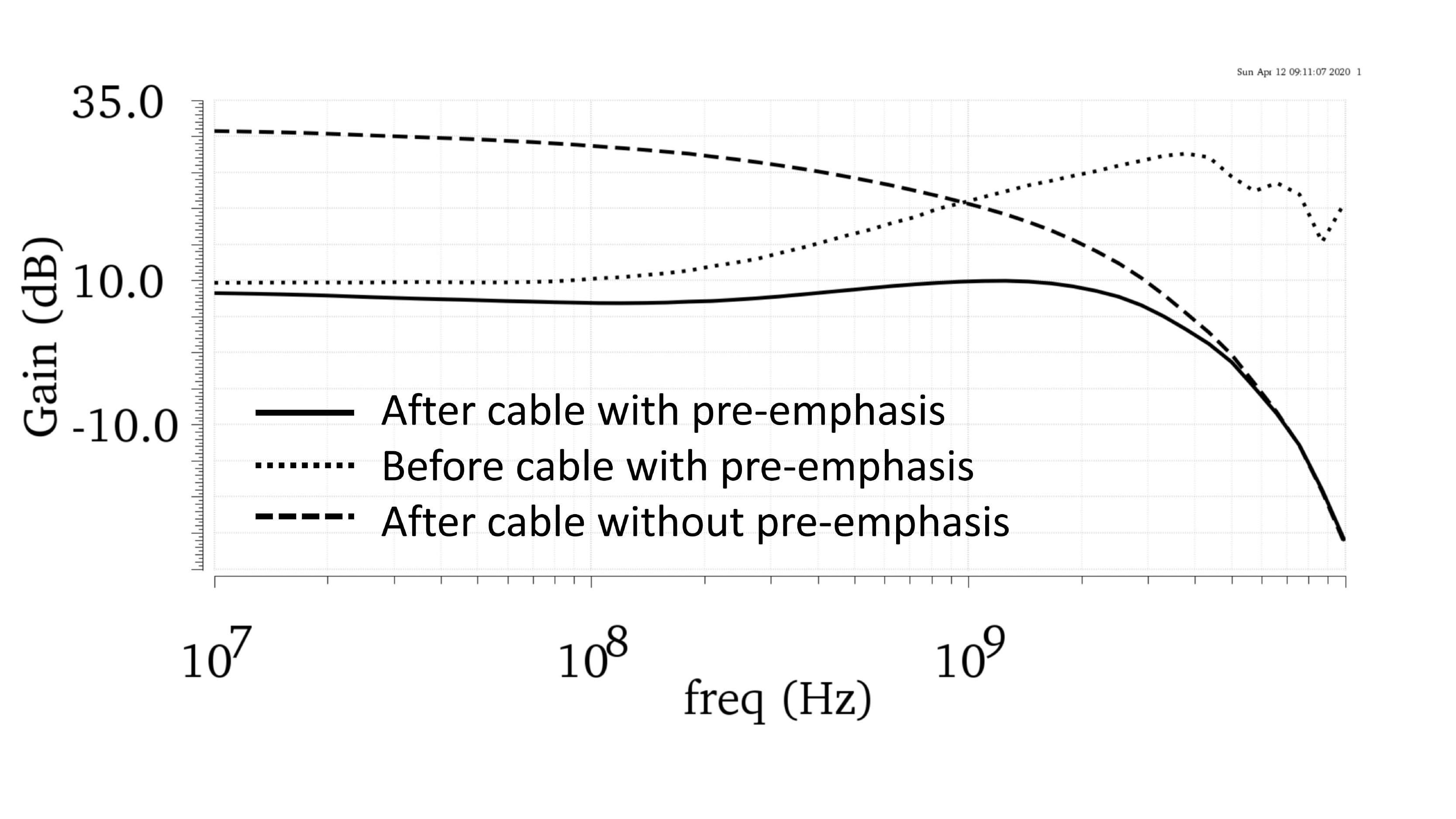}
        	\caption{AC response of downstream channel}
        	\label{fig4:Downstream_AC}
        \end{figure}

        A 2.56 Gbps data from VTRx+ is transmitted to the front-end module through the same electrical cable as the upstream channel. The downstream channel in GBCR1 is designed to pre-emphasize these signals before sending them through the cable which causes a 11.7 dB loss at the Nyquist frequency of 1.28 GHz. The pre-emphasizer stage \cite{Ref13}, which has the CTLE structure is employed with up to 13 dB peaking gain. The gain of the passive attenuator is -3.5 dB to attenuate the maximum amplitude signals from the former LA stages to avoid the saturation of the signal. The AC response of downstream channel is shown in Figure \ref{fig4:Downstream_AC}.
       
    \subsection{GBCR1 Test Results}
\begin{figure}[h]
            \centering
            \setlength{\abovecaptionskip}{0pt}
            \includegraphics[width=1\linewidth]{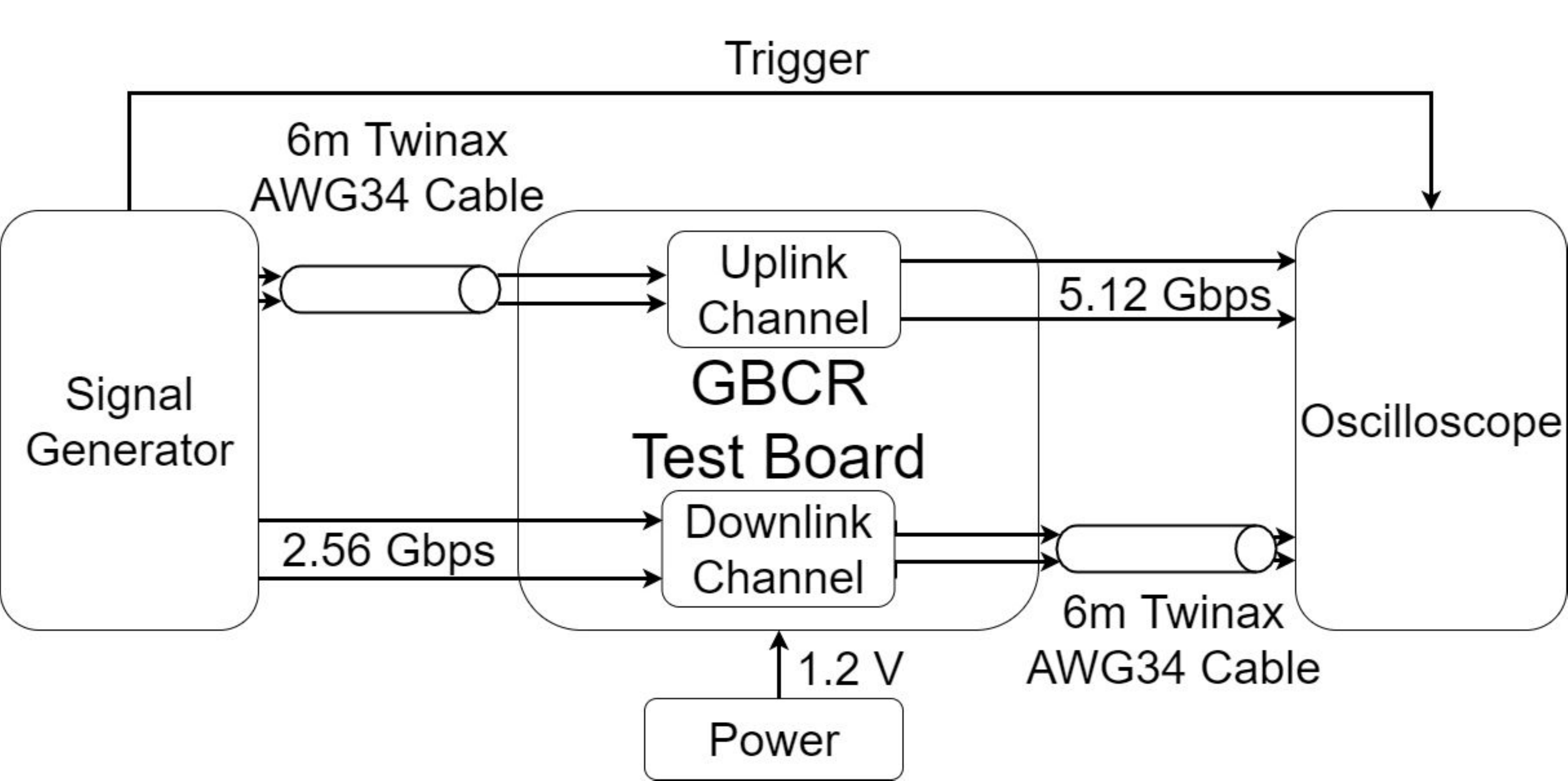}
            \caption{Block Diagram of GBCR1 Test Bench}
            \label{fig5:GBCR1_Testbench}
        \end{figure}

      The test block diagram and the actual test bench are in Figure \ref{fig5:GBCR1_Testbench}. In the test we used the only 6 meter AWG34 Twinax cable that we had. We have updated the simulation with this 6 meter cable in order to compare it with the test results.

\begin{figure}[h]
            \centering
            \setlength{\abovecaptionskip}{0pt}
            \includegraphics[width=1\linewidth]{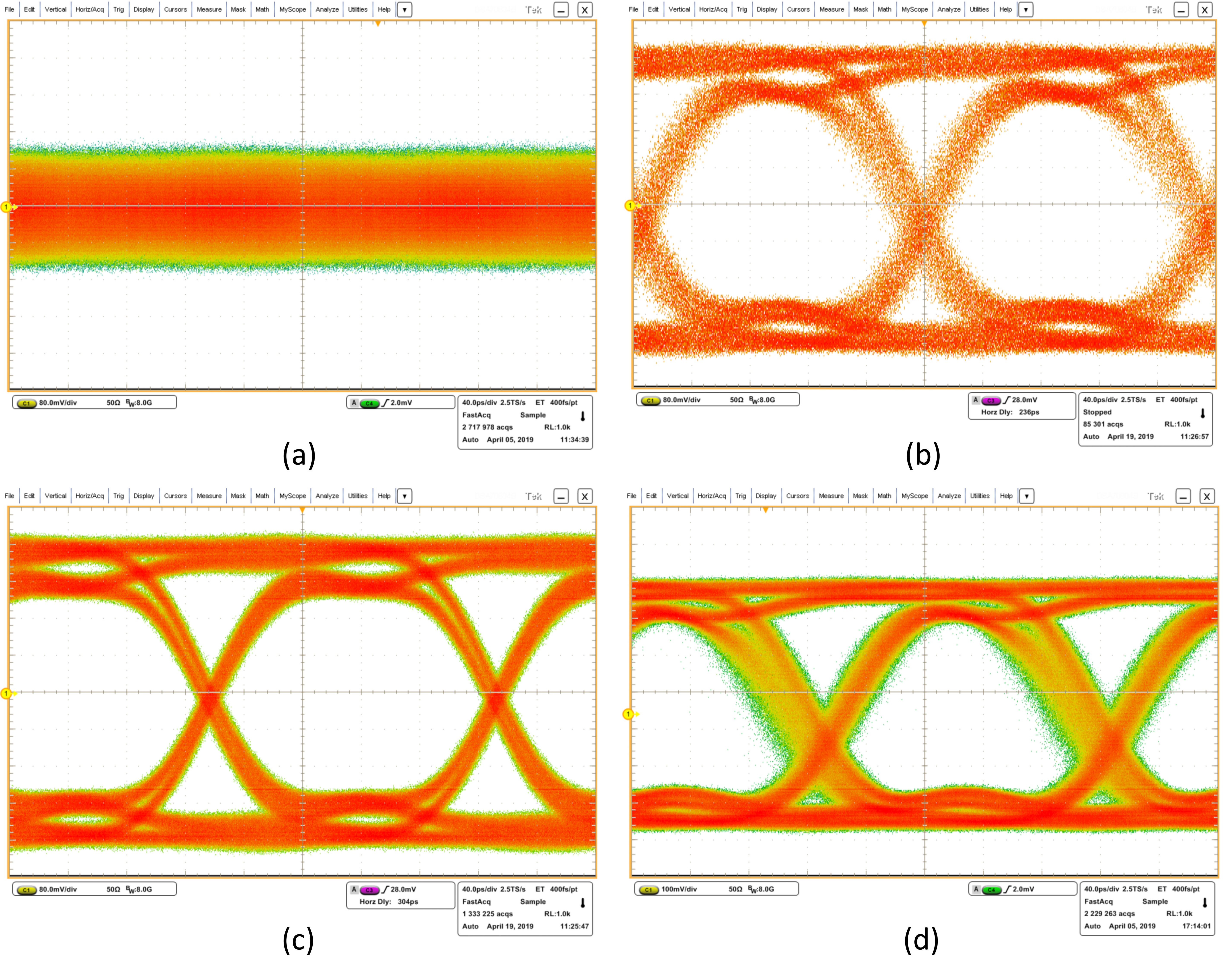}
            \caption{Eye diagrams (a) after the 6m Twinax cable and before the GBCR, (b) after the GBCR (CDR is off) (c), after the GBCR (CDR is on) (d) after the GBCR, one example with obvious DC offset}
            \label{fig6:Eye_diagrams}
        \end{figure}

        Figure \ref{fig6:Eye_diagrams} shows the eye diagrams at different testing points of an upstream channel. As shown in Figure \ref{fig6:Eye_diagrams}(a), when the data from signal generator passes through the 6 meter cable, the eye is completely closed. The eye opens again after being recovered by the equalizer. When the CDR is off and the retiming is bypassed, the peak-peak jitter is 79 ps and the random jitter is 4.3 ps (RMS) under the resistance setting of 0 and the capacitance setting of 3 of the equalizers. When the CDR is turned on, the retiming circuit reduces the peak-peak jitter to 35 ps and the random jitter 1.5 ps (RMS). The CDR significantly reduces the jitter at the cost of 44 mW, a 61\% increase of the channel power.
        
        We also observed some significant DC offset in the output eye diagram as shown in Figure \ref{fig6:Eye_diagrams}(d). We believe the DC offset is from the mismatch in the chip and we confirm that in the simulation. To suppress the jitter from the DC offset, a DC offset cancellation circuit is employed in GBCR2.

\begin{figure}[h]
            \centering
            \setlength{\abovecaptionskip}{0pt}
            \includegraphics[width=1\linewidth]{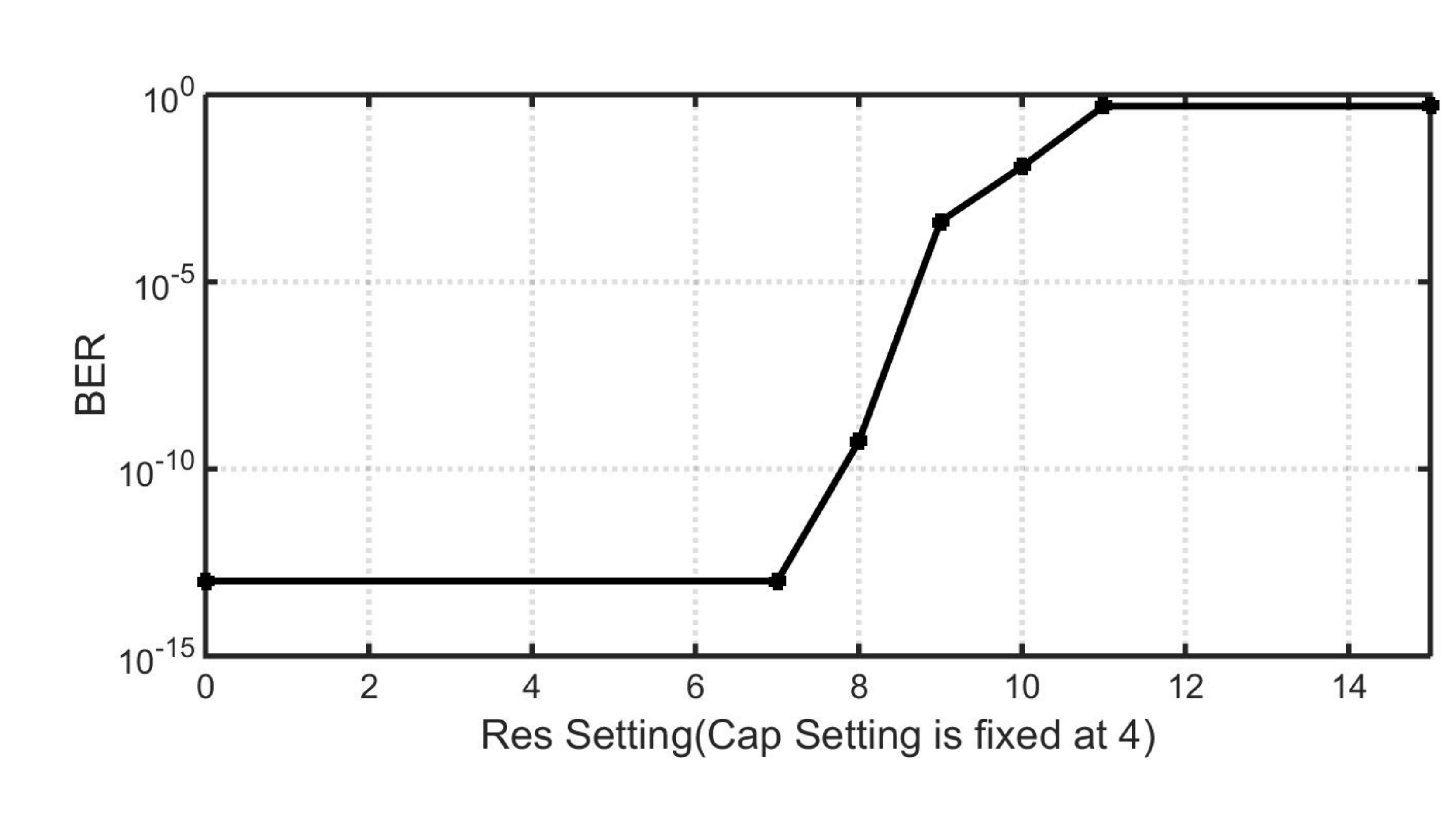}
            \caption{BER Test Results}
            \label{fig7:BER_Test_Results}
        \end{figure}

        Figure \ref{fig7:BER_Test_Results} shows the results of bit error rate (BER), which are obtained by scanning the resistance setting as the capacitance setting is fixed at 4. When the value of resistance setting in CTLEs is from 0 to 7, there are 0 errors during the 20 minute test. The BER is estimated to be below 3.7*$10^{-13}$ with 90\% confidence level which meets the specification of 1*$10^{-12}$. As the resistance setting grows, the BER increases. These results indicate that, when the equalizers in the upstream channels are properly configured, GBCR1 can recover the data correctly.  

\begin{figure}[h]
            \centering
            \setlength{\abovecaptionskip}{0pt}
            \includegraphics[width=1\linewidth]{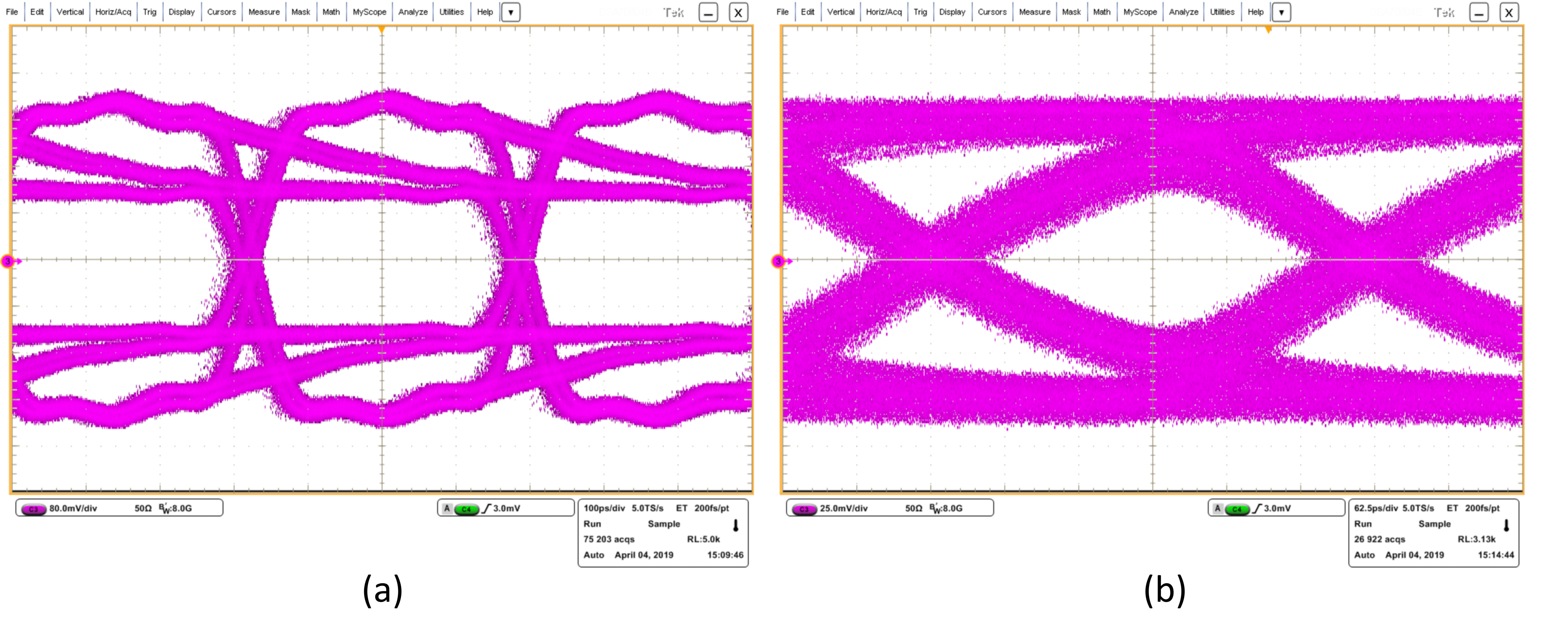}
            \caption{Eye diagrams before (a) and after the Twinax cable (b)}
            \label{fig8:Eye_diagrams}
        \end{figure}

        The downstream channel has been tested with the same cable. Figure \ref{fig8:Eye_diagrams} shows the eye diagrams of the downstream channel before and after the Twinax cable. The overshoot, as shown in Figure \ref{fig8:Eye_diagrams}(a), in the eye diagram before the cable, is obvious because of the pre-emphasis. After the cable, the eye diagram, as shown in Figure \ref{fig8:Eye_diagrams}(b), is still open. The measured peak-peak jitter is 138 ps and the random jitter is 4.1 ps (RMS), consistent with the simulation results of the 6 meter cable.
        
        GBCR1 has also been irradiated with gamma from a Co-60 source. It survives the total ionizing dose up to 200 kGy which is the maximum required dose. The chip was kept powered on during the entire irradiation. We did not observe significant degradation in the amplitude. The jitter of upstream channel changed no more than 2 ps after the radiation in both cases of CDR on and off. 

\section{GBCR2}
\begin{figure}[h]
        \centering
        \setlength{\abovecaptionskip}{0pt}
        \includegraphics[width=1\linewidth]{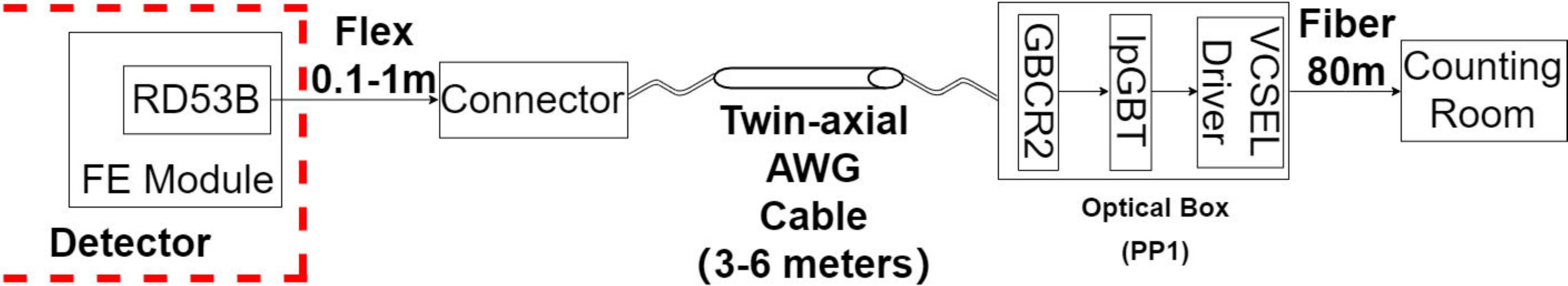}
        \caption{Updated ITk Readout Plan}
        \label{fig9:Updated_ITk}
    \end{figure}

   In the baseline of the data transmission scheme, as shown in Figure \ref{fig9:Updated_ITk}, the sensor readout ASIC RD53B \cite{Ref14} sends data through the FLEX cable and the AWG34 Twinax cable with various length, at a data rate of 1.28 Gbps per channel. The FLEX ranges from 0.1 to 1 meter and the Twinax from 3 to 6 meters. This leads to different high frequency loss. To cope with it, the equalization strength in GBCR2 needs to be programmable. 

\begin{figure}[h]
        \centering
        \setlength{\abovecaptionskip}{0pt}
        \includegraphics[width=1\linewidth]{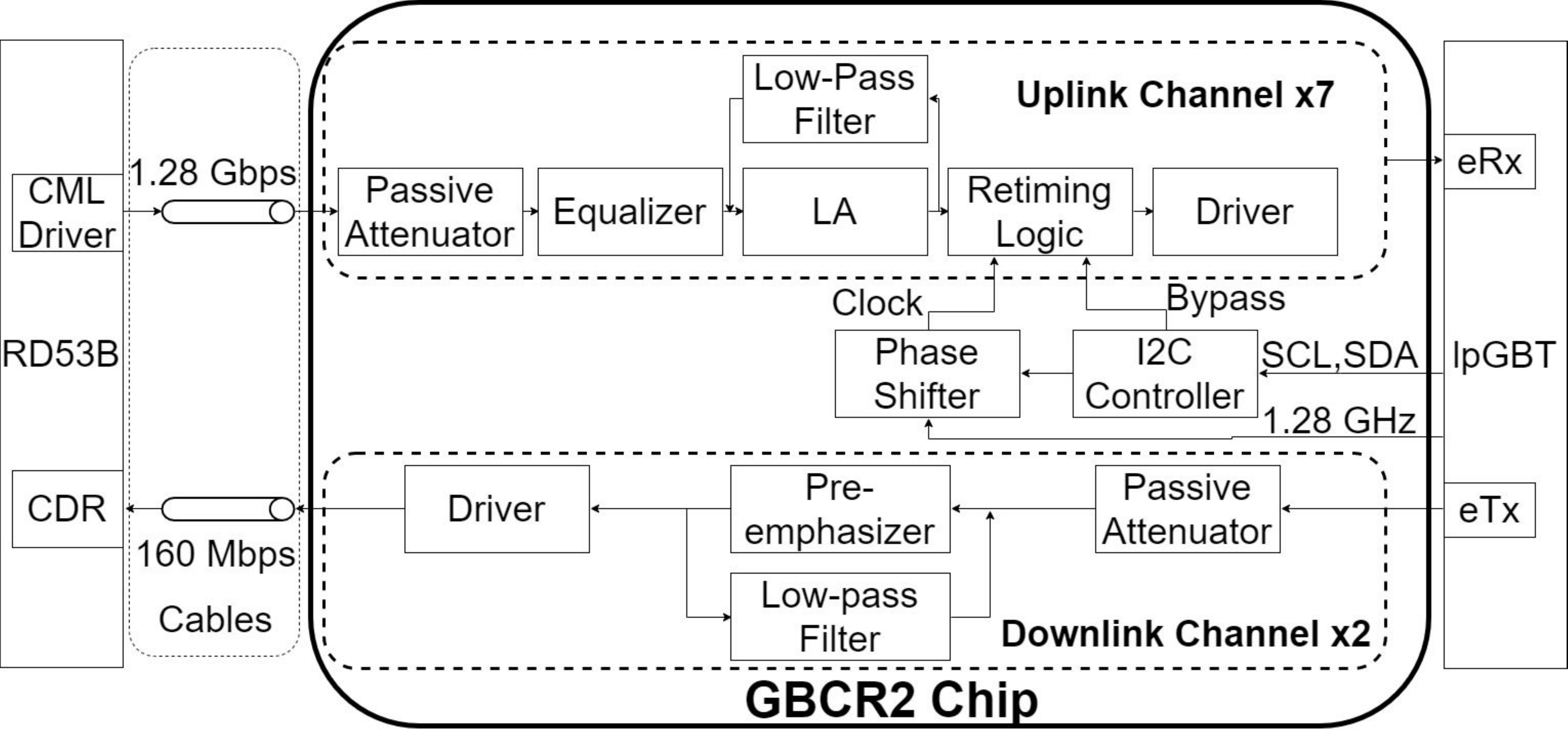}
        \caption{The Block Diagram of GBCR2}
        \label{fig10:GBCR2}
    \end{figure}

   Figure \ref{fig10:GBCR2} shows the block diagram of GBCR2. There are seven uplink channels and two downlink channels. Compared with GBCR1, the CDR module is removed in this design because it consumes 36.4 mA each which is too much for the power budget in the new design of 7 channels. Instead, an external clock of 1.28 GHz with a phase shifter is used to achieve the retiming function. All these uplink channels share the same phase shifter \cite{Ref15}\cite{Ref16} which is adapted from lpGBT. An I2C slave provides configure and control signals for all the channels in GBCR2.

    \subsection{Uplink Channel}
        As seen in Figure \ref{fig10:GBCR2}, the uplink channel consists of a passive attenuator, an equalizer, retiming logic, DC offset cancellation circuits and CML driver. The passive attenuator is programmable to attenuate the signal from RD53B to avoid the saturation distortion in case the signal amplitude is large. The equalizer has a high-pass differential stage to compensate the signal loss. The DC cancellation circuit is designed to reduce the DC offset caused by mismatch in the amplifiers. The retiming circuit of each channel shares one phase shifter but the sampling clock phase of each channel is programmable independently. A CML driver is used to send the recovered data to the eRx module in lpGBT.
        
\begin{figure}[h]
            \centering
            \setlength{\abovecaptionskip}{0pt}
            \includegraphics[width=1\linewidth]{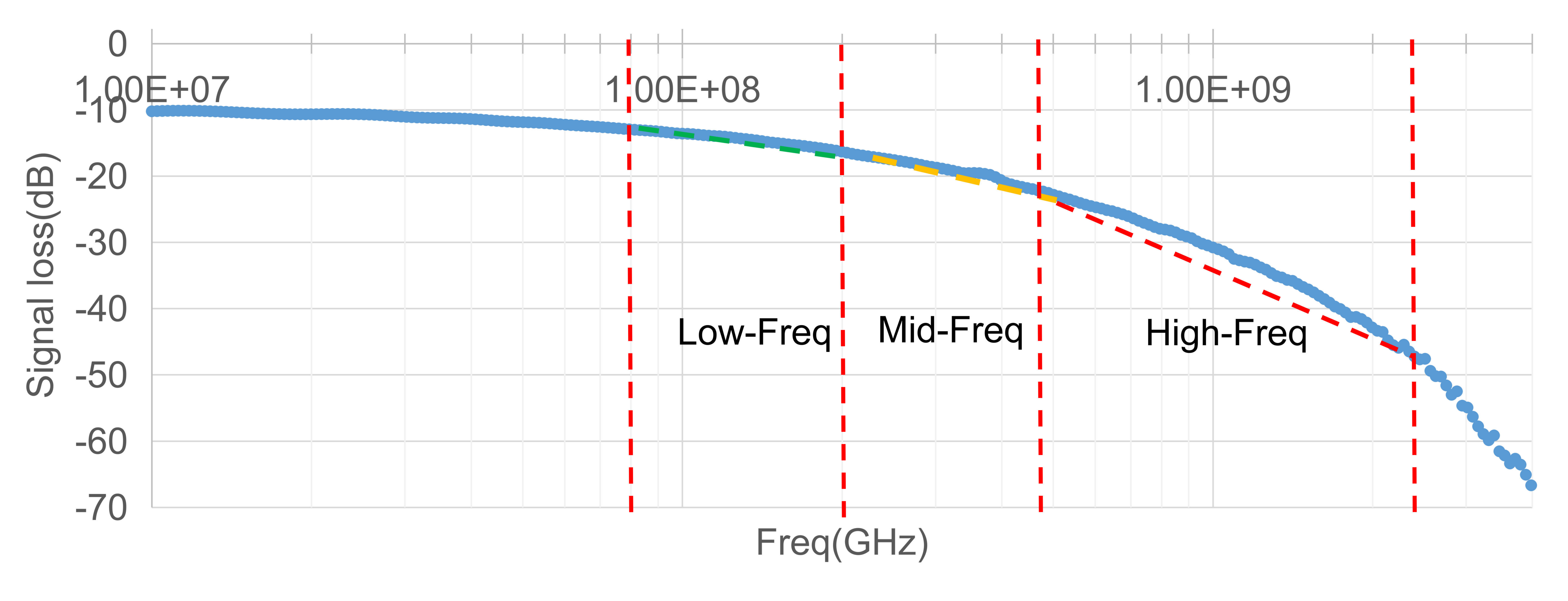}
            \caption{Simulation of Cable Loss based on 1-m flex model and 6-m Twinax model in series}
            \label{fig11:cable_loss_analyse}
        \end{figure}

        From the frequency response curve of the signal loss in the worst case (1 meter FLEX cable and 6 meter Twinax cable), we divided the curve into 3 ranges, as shown in Figure \ref{fig11:cable_loss_analyse}. In the high frequency range from 0.4-1.3 GHz, the decay slope is 27.3 dB/dec. In the middle frequency range from 0.2 to 0.4 GHz, it is about 14.3 dB/dec, which is about half of the high-frequency range. In the low frequency range from 0.08 to 0.2 GHz, the decay slope is 8.5 dB/dec, about one third of the high frequency range. In this case, the multiple stages of CTLE are adopted to compensate the loss due to the long-distance transmission. But for the high frequency part, the slope is too steep to be compensated with one order zero. We design a 3 identical CTLE stages to compensate with 3 order zero for high frequency range. For the medium and low frequency parts, each of them just employs one CTLE stage to recover the signal loss in their range respectively. The entire equalizer stage can provide 30 dB peaking strength at most. As the CTLE in GBCR1, the resistance of Rs in GBCR2 is programmable by 4 control bits to tune the zero position and the equalization. But the capacitance of Cs is constant in order to keep up the pole position.

\begin{figure}[h]
            \centering
            \setlength{\abovecaptionskip}{0pt}
            \includegraphics[width=1\linewidth]{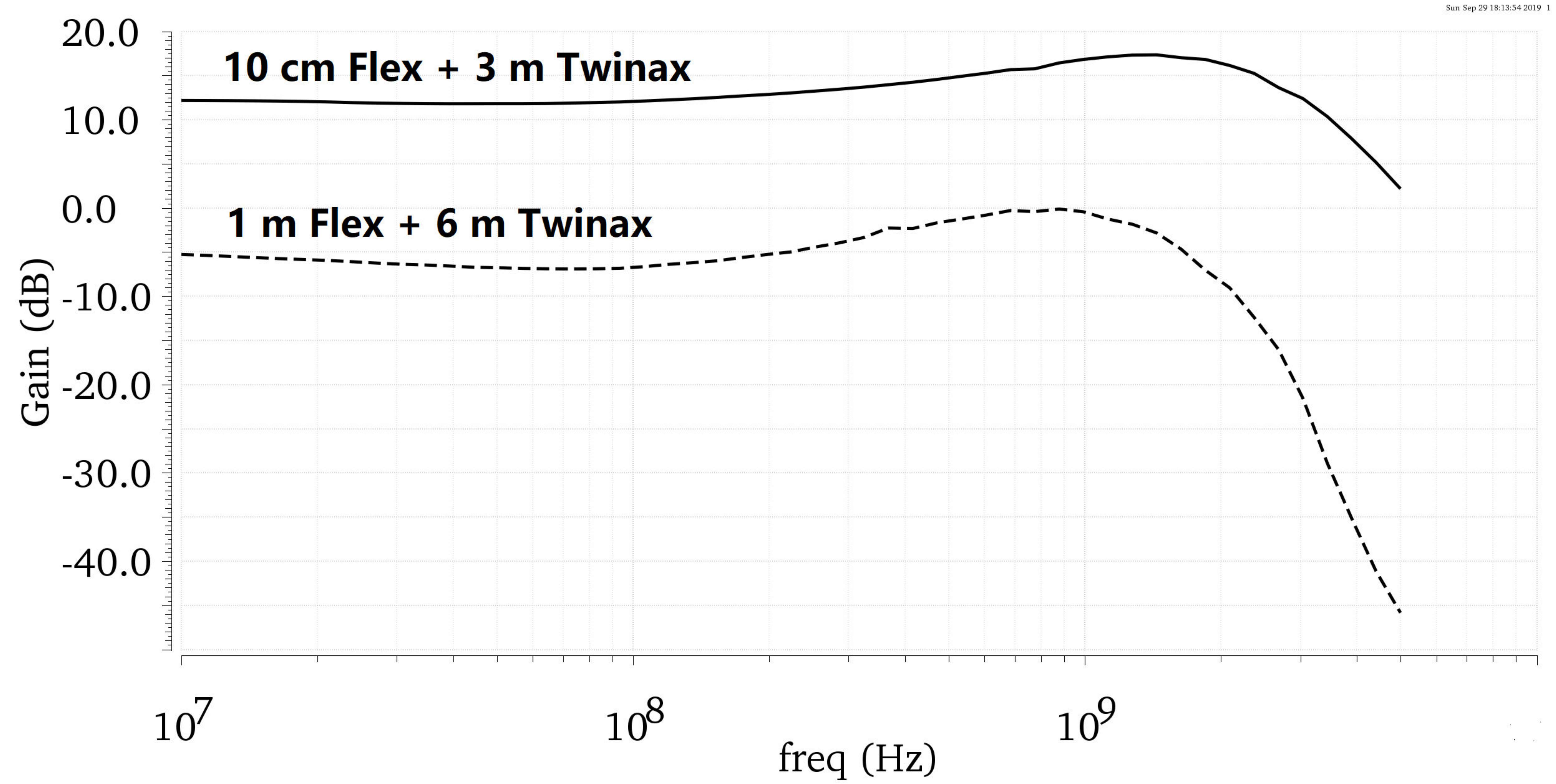}
            \caption{AC Response of the recovered signals}
            \label{fig12:AC_Response}
        \end{figure}

        In the multiple-stage structure, we keep the DC gain of each stage around 0 dB and amplify the high-frequency signal only. After recovered by the equalizers in uplink channel, the frequency response of the equalizer output is shown in Figure \ref{fig12:AC_Response}. The gain is 12.2 dB when using the shortest cable and it will be -5.2 dB in the situation of the longest cable. 

\begin{figure}[h]
            \centering
            \setlength{\abovecaptionskip}{0pt}
            \includegraphics[width=1\linewidth]{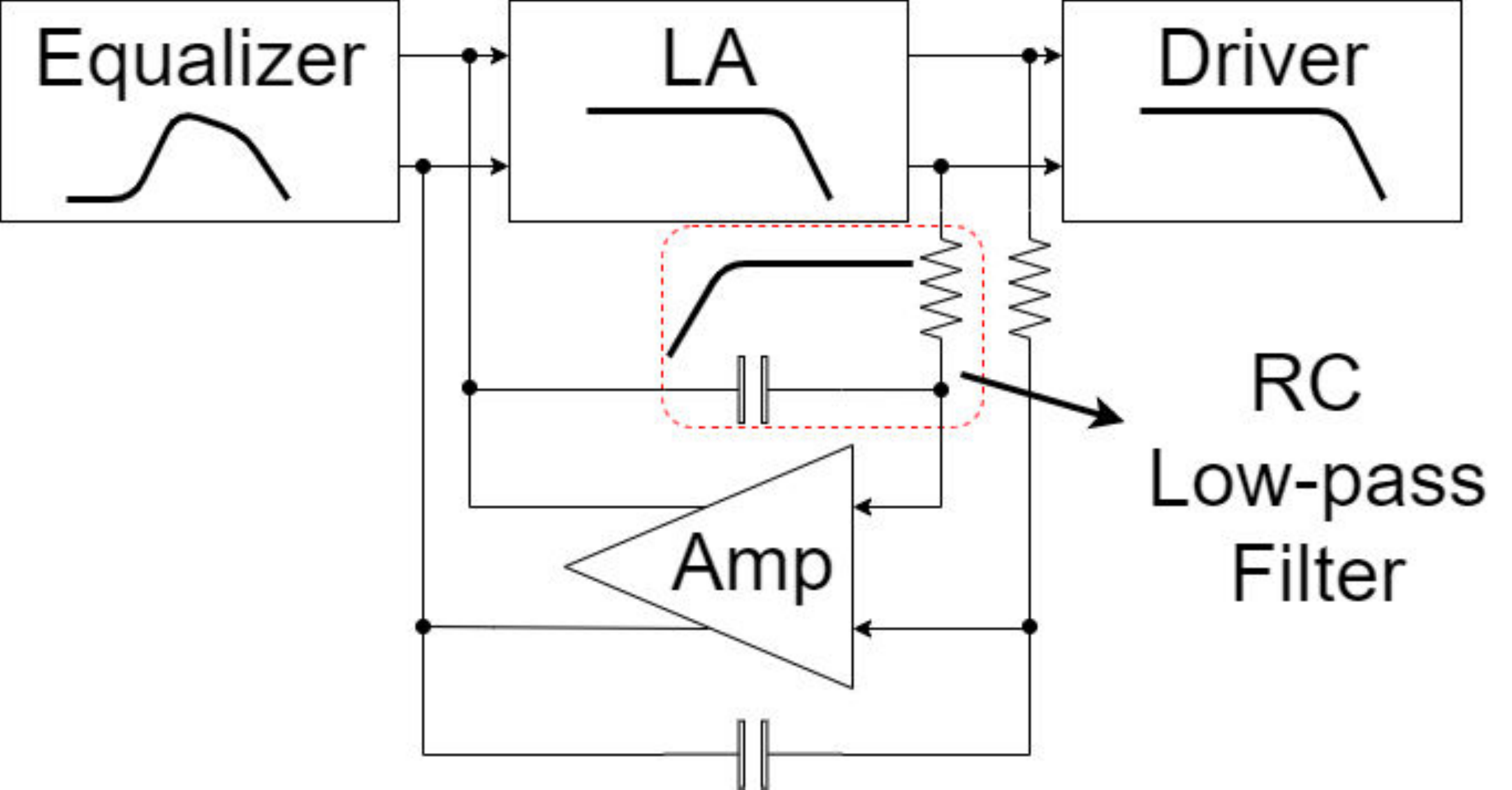}
            \caption{DC-offset Cancellation Circuit}
            \label{fig13:Cancellation_Circuit}
        \end{figure}

        As discussed in Section 2, a DC offset cancellation circuit is needed to suppress the DC offset from mismatch. The circuit is shown in Figure \ref{fig13:Cancellation_Circuit}. The high frequency component is removed by a low pass RC filter and the DC offset is amplified and fed back to the input stage of LA to adjust and cancel the DC offset. The DC offset cancellation circuit causes a low cut-off frequency in the main signal path. The low cut-off frequency is 115 kHz, which has no effect on the data encoded in 64b/66b line code. Because the gain of the feedback amplifier is 100 dB, we expect that the DC offset is highly suppressed.

        In the GBCR1 design, the CML driver operates at 5.12 Gbps without the possibility to adjust the output amplitude. In order to save power, a simple circuit to tune the output amplitude is implemented. Three switches in the bias circuit are used to control the tail current for the CML driver and its amplitude consequently. The range of the output amplitude is from 44.6 mV to 313.7 mV for single-ended with each step of about 40 mV. The power consumption of this driver is from 1.8 mA to 12 mA. 
        
\begin{figure}[h]
              \centering
              \setlength{\abovecaptionskip}{0pt}
              \includegraphics[width=1\linewidth]{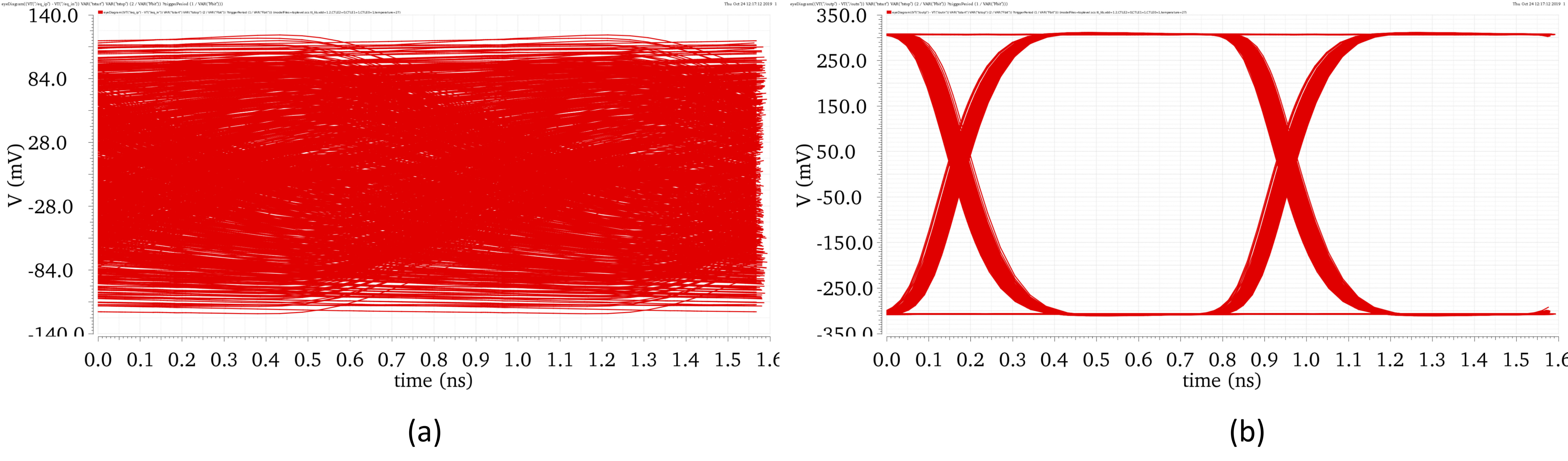}
              \caption{Simulation results of before (a) and after recovered by the uplink channel (b)}
              \label{fig14:Simulation_results}
          \end{figure}

        For the uplink channel, with the longest cable length (1 meter FLEX and 6 meter Twinax), when the DC offset cancellation circuit is off, the jitter of final output of the uplink channel is below 100 ps in simulations for all the process corners. The typical eye diagram is shown in Figure \ref{fig14:Simulation_results}, in which the jitter is about 65.9 ps.

    \subsection{Downlink Channel}
        The downlink channel, which works at 160 Mbps, is a part of the command link, which includes the eTx in lpGBT, the downlink channel in GBCR2, and the CDR in RD53B. It is important to suppress the ISI jitter of the input signal of the CDR in RD53B because the recovered clock signal is also used for high-speed data transmission up to 1.28 Gbps. The GBCR2 downlink channel receives the 160 Mbps data from lpGBT, pre-emphasize them and then transmit the data to RD53B through the same cables as the ones used in uplink channels. As shown in Figure \ref{fig10:GBCR2}, this channel also consists of a programmable passive attenuator, a pre-emphasizer, a low-pass filter circuit and a CML output driver. The pre-emphasizer has two stages of CTLE which is the same structure as used in the uplink channel. But because the data rate is much lower, the parameters in the downlink channel are redesigned. Its zero is at 40 MHz and the pole is at 120 MHz. There is also a 3-bit control in the pre-emphasizer stage to adjust the peaking strength which is up to 14.8 dB. 
        
\begin{figure}[h]
            \centering
            \setlength{\abovecaptionskip}{0pt}
            \includegraphics[width=1\linewidth]{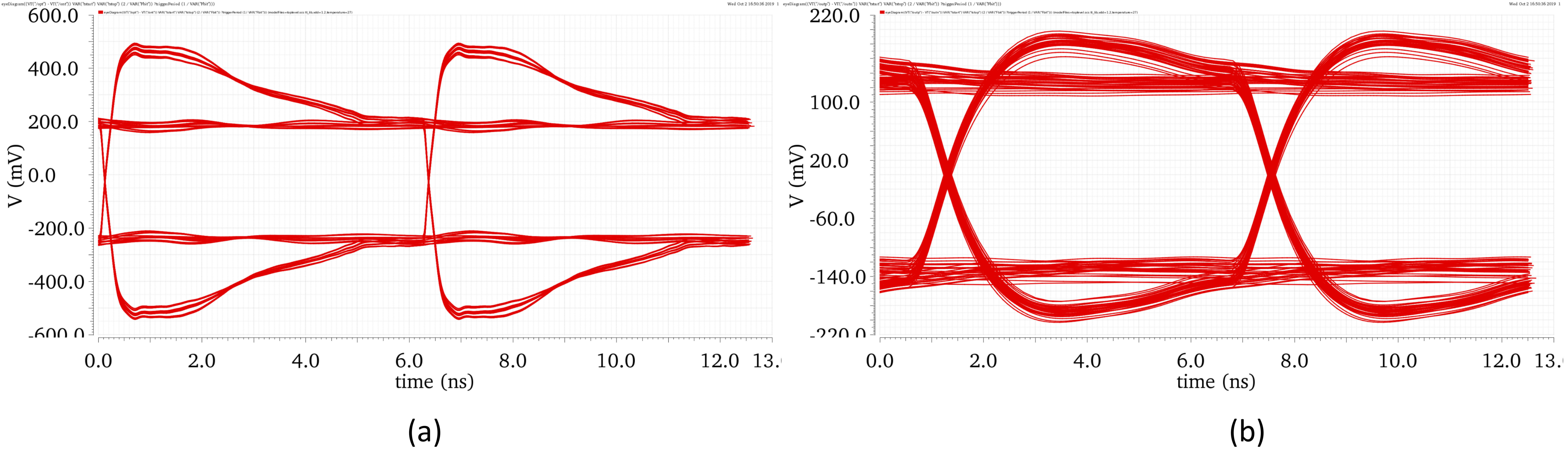}
            \caption{Simulation results of before (a) and after (b) the cables of the downlink channel}
            \label{fig15:Simulation_Results}
        \end{figure}

        Even in the transmission situation with the longest cables, the downstream channel significantly reduce the ISI jitter from 500 ps to 157 ps. The eye diagrams before and after the cables are shown in Figure \ref{fig15:Simulation_Results}.

        As a conclusion, based on simulation GBCR2 will work well to receive and transmit the specific data by the uplink and downlink channels. The table below are the details for the performance of this chip. In the Equalizer mode, where the retiming circuit and phase shifter are off to save the power, the total power consumption is 79 mW. In the retiming mode, the power consumption increases to 105 mW but ISI jitter is further suppressed. GBCR2 was submitted in November 2019 and will be tested in 2020.

		\begin{table*}[ht]
			\tiny
			\caption{The jitter performance of GBCR2}
			\scalebox{1.08}{
			\begin{tabular}{|p{0.1\textwidth}<{\centering}|p{0.1\textwidth}<{\centering}|p{0.2\textwidth}<{\centering}|p{0.2\textwidth}<{\centering}|p{0.2\textwidth}<{\centering}|}
				\hline
				& Mode              & \multicolumn{1}{l|}{Input Jitter from RD53 (ps)} & Jitter from cable + GBCR (ps) & \begin{tabular}[c]{@{}l@{}}Estimated Output jitter to\\   lpGBT (ps)\end{tabular} \\ \hline
				\multirow{2}{*}{Uplink channel} & Rx Equalizer mode & \multirow{2}{*}{100}                             & 81                            & 181                                                                               \\ \cline{2-2} \cline{4-5}
				& Rx Retiming mode  &                                                  & 11                            & 80                                                                                \\ \hline
				\multicolumn{5}{|c|}{The jitter of downlink output is 156.9 ps.}                                                                                                                                                    \\ \hline
			\end{tabular}
		}
		\end{table*}

\section{Summary}
    In this paper, we present two prototypes of a transceiver ASIC, GBCR1 and GBCR2, in a 65-nm CMOS technology for the ATLAS ITk detector readout upgrade. GBCR1 has been tested and GBCR2 is submitted but not tested yet. 
    
    The uplink channel of GBCR1, which operates at 5.12 Gbps, recovers the data through a 6 meter AWG34 Twinax cable. The ISI jitter is 79.9 ps when the CDR is off and the jitter decreases to 35.5 ps when the CDR is on. The downlink transmits the 2.56 Gbps data from lpGBT well in the test. The total power consumption of GBCR1 is 192 mW. The chip survives the total ionizing dose of 200 kGy with a gamma source.
    
    We redesigned the equalizer in the GBCR2 to adapt to the new readout baseline. By using an external clock with a shared phase-shifter instead of a CDR for each channel for the re-timing, the power consumption of each channel is kept to 105 mW. The output signal jitter is to 70 ps, which meets the requirement of the eRx in lpGBT. The downlink channel is optimized for 160 Mbps data rate and suppresses the ISI jitter for the transmission in the cables. We expect to test GBCR2 soon.

\section*{Acknowledgments}
    This work was supported by the US-ATLAS phase-2 upgrade grant administrated by the US-ATLAS phase-2 upgrade project office and the Office of High Energy Physics of the U.S. Department of Energy under contract DE-AC02-05CH11231. The authors would like to be grateful to Andrew Young and Dong Su from SLAC, Maurice Garcia-Sciveres and Veronica Wallangen from LBNL, James Kierstead from BNL for their kind help on the test of GBCR1.

\end{document}